\documentclass[twocolumn]{IEEEtran}
\pdfoutput=1
\usepackage{graphicx}
\usepackage{color}
\usepackage{SIunits}
\usepackage{multirow}
\usepackage[pdfborder=0]{hyperref}

\begin{document}
\newcommand{\rtHz}{$\sqrt{\mbox{Hz}}$}
\newcommand{\mycomment}[1]{$\dagger$ \marginpar{\it\tiny $\dagger$ #1}} % used for margin comments

\title{A Biasing and Demodulation System\\
for Kilopixel TES Bolometer Arrays}

\author{
	Graeme~Smecher,~\IEEEmembership{Member,~IEEE,}
	Fran\c{c}ois Aubin,
	Eric~Bissonnette,
	Matt~Dobbs,
	Peter~Hyland,
	and~Kevin~MacDermid
\IEEEcompsocitemizethanks{\IEEEcompsocthanksitem G. Smecher, F. Aubin, E.
	Bissonnette, M. Dobbs, P. Hyland, and K. Macdermid are with the
	University of McGill, Montr\'eal, Qu\'ebec, Canada.
	\protect\\
	\protect\\
%This paper has been submitted for review to the IEEE Trans. on
%Instrumentation and Measurement. If accepted for publication, the copy of
%record will be available at the IEEE Digital Library.
%	\protect\\
%	\protect\\
	E-mail: graeme.smecher@mail.mcgill.ca}
}

\markboth{IEEE Trans.\ on Instrum.\ Meas.}{Smecher \MakeLowercase{\textit{et
al}}: Biasing and Demodulation for Kilopixel TES Bolometer Arrays}

\maketitle

\begin{abstract}
  We describe the signal-processing logic, firmware, and software for a
  frequency-domain multiplexed (FDM) biasing and demodulation system that
  reads out Transition Edge Sensor (TES) bolometer arrays for mm-wavelength
  cosmology telescopes.  This system replaces a mixed-signal readout backend
  with a much smaller, more power-efficient system relying on
  Field-Programmable Gate Arrays (FPGAs) for control, computation and signal
  processing. The new system is sufficiently robust, automated, and power
  efficient to be flown on stratospheric balloon-borne telescopes and is being
  further developed for satellite applications.
\end{abstract}


%\bstctlcite{IEEEexample:BSTcontrol}

\section{Introduction}

Large arrays of Transition Edge Sensor (TES) bolometers have recently
seen widespread application to dark matter detection experiments and
astronomical telescopes in the millimeter, sub-millimeter,
far-infrared, soft x-ray, and gamma-ray bands (see, for example,
\cite{Cabrera2000,apexSZ_instrument,Grainger2008,Carlstrom2009,
Holland2006,deKorte2008,Mauskopf2008,Doriese2007}). This
is due to their exquisite sensitivity for incoherent detection, ease
for fabrication of large monolithic arrays using photolithographic
techniques, and relative insensitivity to microphonics due to their
low impedance. The detector arrays typically operate at sub-Kelvin
temperatures, so the thermal load presented by the wires connecting
cryogenic hardware to room-temperature electronics is a significant design
constraint.  In such situations, it is desirable to multiplex many signals
onto a reduced number of wires, using, for example, time-domain
multiplexing~\cite{Irwin2002} (TDM) or frequency-domain
multiplexing (FDM)~\cite{Lanting2005,Dobbs2008}.

In this paper, we describe the logic and firmware design of the Digital
Frequency-domain Multiplexer (DFMUX), a power- and space-efficient FDM
system that multiplexes up to 16 TES bolometers on a single readout
module consisting of one set of wires entering the cryostat. Each
module uses a single series array Superconducting Quantum Interference
Device (SQUID)~\cite{NIST-arrays} operating at $4 \degree \kelvin$ as a
transimpedance pre-amplifier.  This readout system is being used for
several mm-wavelength telescopes including the EBEX Balloon-borne
polarimeter~\cite{Grainger2008} and POLARBEAR
experiment~\cite{Lee2008}.

The FDM modulator and demodulator logic for each readout module is
implemented on a Xilinx Virtex-4 FPGA~\cite{Virtex4}. In addition to a high
channel density on each set of wires, this system achieves a high overall
density, supporting 4 multiplexed readout modules on a single 6U VME circuit
board with one FPGA; up to 20 such cards have been operated in a single VME
subrack. Where multiple subracks are used, clock and timing signals are
daisy-chained between them.

The DFMUX is intended to supersede a mixed-signal system described
in~\cite{Lanting2005,Dobbs2006,Carlstrom2009} with a system that is
sufficiently small and power-efficient for balloon-borne experiments.  An
overview may be found in~\cite{Dobbs2008}, with a description of an
early system design. The high-level software control of the system,
along with its use to tune the cryogenic electronics, is documented
in~\cite{Macdermid2009}.  In this paper, we focus on a system-level
description and analysis of the DFMUX signal path, and highlight recent
changes that extend the design, improve channel density, and increase
performance in the noise environment exhibited in the field. We begin by
describing the system's basic operation in
Section~\ref{sec:system-description}.  We focus on the digital signal path in
Section~\ref{sec:signalpath}. We then examine the firmware and on-board
software in Section~\ref{sec:software}. We explore the system performance in
Section~\ref{sec:performance} and conclude with
Section~\ref{sec:conclusions}.

\section{System Description\label{sec:system-description}}

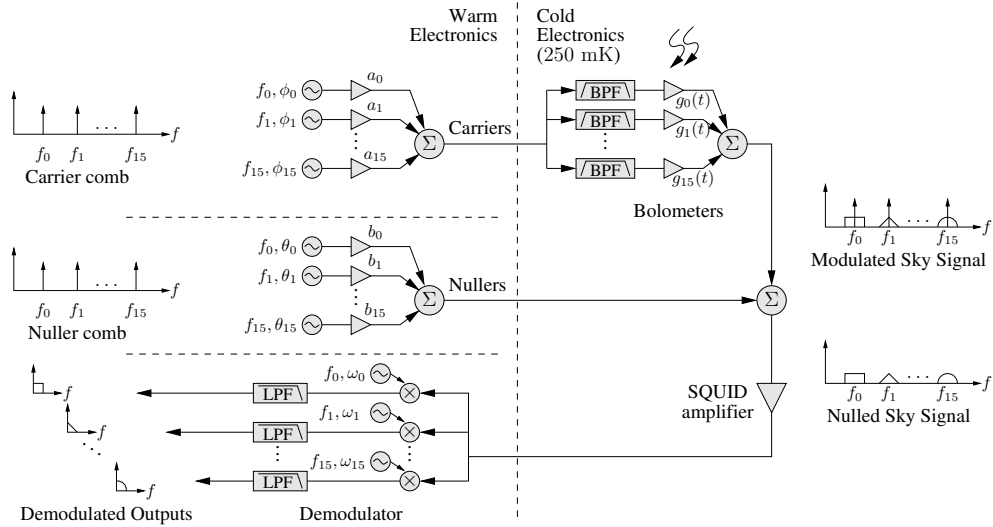
\begin{figure*}
\centerline{\scalebox{0.65}{
	\begin{picture}(0,0)%
	\includegraphics{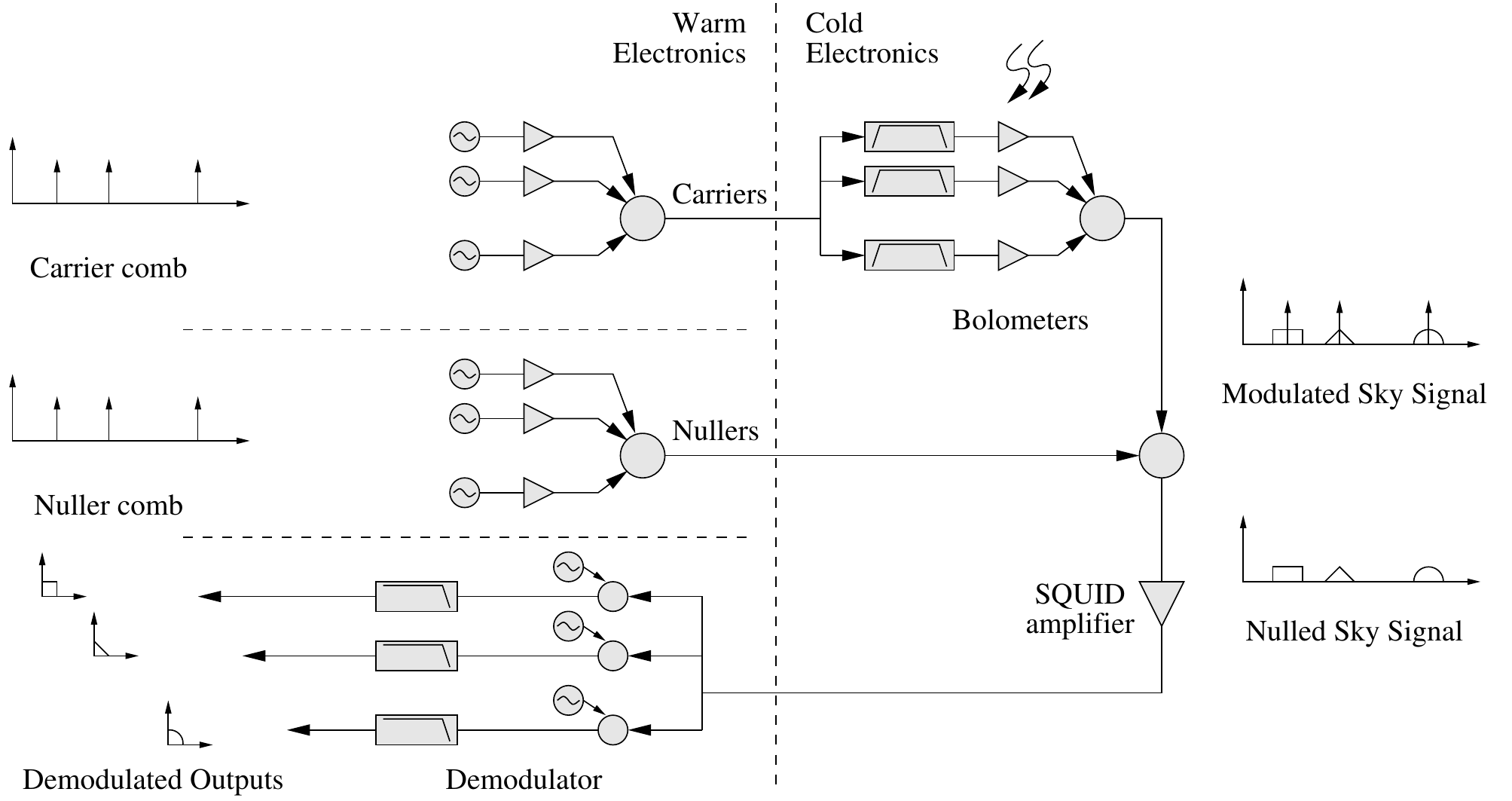}%
	\end{picture}%
	\setlength{\unitlength}{4144sp}%
	\begingroup\makeatletter\ifx\SetFigFont\undefined%
	\gdef\SetFigFont#1#2#3#4#5{%
	  \normalfont\fontsize{#1}{#2pt}%
	  \fontfamily{#3}\fontseries{#4}\fontshape{#5}%
	  \selectfont}%
	\fi\endgroup%
	\begin{picture}(9173,4851)(-6149,-2740)
	\put(2251, 74){\makebox(0,0)[b]{\smash{{\SetFigFont{12}{14.4}{\rmdefault}{\mddefault}{\updefault}{\color[rgb]{0,0,0}$\ldots$}}}}}
	\put(1666,-106){\makebox(0,0)[b]{\smash{{\SetFigFont{10}{12.0}{\rmdefault}{\mddefault}{\updefault}{\color[rgb]{0,0,0}$f_0$}}}}}
	\put(1981,-106){\makebox(0,0)[b]{\smash{{\SetFigFont{10}{12.0}{\rmdefault}{\mddefault}{\updefault}{\color[rgb]{0,0,0}$f_1$}}}}}
	\put(2521,-106){\makebox(0,0)[b]{\smash{{\SetFigFont{10}{12.0}{\rmdefault}{\mddefault}{\updefault}{\color[rgb]{0,0,0}$f_{15}$}}}}}
	\put(-5219,-511){\makebox(0,0)[b]{\smash{{\SetFigFont{12}{14.4}{\rmdefault}{\mddefault}{\updefault}{\color[rgb]{0,0,0}$\ldots$}}}}}
	\put(-4949,-781){\makebox(0,0)[b]{\smash{{\SetFigFont{10}{12.0}{\rmdefault}{\mddefault}{\updefault}{\color[rgb]{0,0,0}$f_{15}$}}}}}
	\put(-5489,-781){\makebox(0,0)[b]{\smash{{\SetFigFont{10}{12.0}{\rmdefault}{\mddefault}{\updefault}{\color[rgb]{0,0,0}$f_1$}}}}}
	\put(-5804,-781){\makebox(0,0)[b]{\smash{{\SetFigFont{10}{12.0}{\rmdefault}{\mddefault}{\updefault}{\color[rgb]{0,0,0}$f_0$}}}}}
	\put(-5219,929){\makebox(0,0)[b]{\smash{{\SetFigFont{12}{14.4}{\rmdefault}{\mddefault}{\updefault}{\color[rgb]{0,0,0}$\ldots$}}}}}
	\put(-4949,659){\makebox(0,0)[b]{\smash{{\SetFigFont{10}{12.0}{\rmdefault}{\mddefault}{\updefault}{\color[rgb]{0,0,0}$f_{15}$}}}}}
	\put(-5489,659){\makebox(0,0)[b]{\smash{{\SetFigFont{10}{12.0}{\rmdefault}{\mddefault}{\updefault}{\color[rgb]{0,0,0}$f_1$}}}}}
	\put(-5804,659){\makebox(0,0)[b]{\smash{{\SetFigFont{10}{12.0}{\rmdefault}{\mddefault}{\updefault}{\color[rgb]{0,0,0}$f_0$}}}}}
	\put(2251,-1366){\makebox(0,0)[b]{\smash{{\SetFigFont{12}{14.4}{\rmdefault}{\mddefault}{\updefault}{\color[rgb]{0,0,0}$\ldots$}}}}}
	\put(1666,-1546){\makebox(0,0)[b]{\smash{{\SetFigFont{10}{12.0}{\rmdefault}{\mddefault}{\updefault}{\color[rgb]{0,0,0}$f_0$}}}}}
	\put(1981,-1546){\makebox(0,0)[b]{\smash{{\SetFigFont{10}{12.0}{\rmdefault}{\mddefault}{\updefault}{\color[rgb]{0,0,0}$f_1$}}}}}
	\put(2521,-1546){\makebox(0,0)[b]{\smash{{\SetFigFont{10}{12.0}{\rmdefault}{\mddefault}{\updefault}{\color[rgb]{0,0,0}$f_{15}$}}}}}
	\put(901,-702){\makebox(0,0)[b]{\smash{{\SetFigFont{12}{14.4}{\rmdefault}{\mddefault}{\updefault}{\color[rgb]{0,0,0}$\Sigma$}}}}}
	\put(-629,749){\makebox(0,0)[b]{\smash{{\SetFigFont{12}{14.4}{\rmdefault}{\mddefault}{\updefault}{\color[rgb]{0,0,0}$\vdots$}}}}}
	\put(541,738){\makebox(0,0)[b]{\smash{{\SetFigFont{12}{14.4}{\rmdefault}{\mddefault}{\updefault}{\color[rgb]{0,0,0}$\Sigma$}}}}}
	\put(181,1154){\makebox(0,0)[b]{\smash{{\SetFigFont{10}{12.0}{\rmdefault}{\mddefault}{\updefault}{\color[rgb]{0,0,0}$g_0(t)$}}}}}
	\put(181,884){\makebox(0,0)[b]{\smash{{\SetFigFont{10}{12.0}{\rmdefault}{\mddefault}{\updefault}{\color[rgb]{0,0,0}$g_1(t)$}}}}}
	\put(181,434){\makebox(0,0)[b]{\smash{{\SetFigFont{10}{12.0}{\rmdefault}{\mddefault}{\updefault}{\color[rgb]{0,0,0}$g_{15}(t)$}}}}}
	\put(-2249,738){\makebox(0,0)[b]{\smash{{\SetFigFont{12}{14.4}{\rmdefault}{\mddefault}{\updefault}{\color[rgb]{0,0,0}$\Sigma$}}}}}
	\put(-2249,-702){\makebox(0,0)[b]{\smash{{\SetFigFont{12}{14.4}{\rmdefault}{\mddefault}{\updefault}{\color[rgb]{0,0,0}$\Sigma$}}}}}
	\put(-2744,1379){\makebox(0,0)[b]{\smash{{\SetFigFont{10}{12.0}{\rmdefault}{\mddefault}{\updefault}{\color[rgb]{0,0,0}$a_0$}}}}}
	\put(-2744,1109){\makebox(0,0)[b]{\smash{{\SetFigFont{10}{12.0}{\rmdefault}{\mddefault}{\updefault}{\color[rgb]{0,0,0}$a_1$}}}}}
	\put(-3464,524){\makebox(0,0)[rb]{\smash{{\SetFigFont{10}{12.0}{\rmdefault}{\mddefault}{\updefault}{\color[rgb]{0,0,0}$f_{15}, \phi_{15}$}}}}}
	\put(-3464,974){\makebox(0,0)[rb]{\smash{{\SetFigFont{10}{12.0}{\rmdefault}{\mddefault}{\updefault}{\color[rgb]{0,0,0}$f_1, \phi_1$}}}}}
	\put(-3464,1244){\makebox(0,0)[rb]{\smash{{\SetFigFont{10}{12.0}{\rmdefault}{\mddefault}{\updefault}{\color[rgb]{0,0,0}$f_0, \phi_0$}}}}}
	\put(-3464,-196){\makebox(0,0)[rb]{\smash{{\SetFigFont{10}{12.0}{\rmdefault}{\mddefault}{\updefault}{\color[rgb]{0,0,0}$f_0, \theta_0$}}}}}
	\put(-3464,-466){\makebox(0,0)[rb]{\smash{{\SetFigFont{10}{12.0}{\rmdefault}{\mddefault}{\updefault}{\color[rgb]{0,0,0}$f_1, \theta_1$}}}}}
	\put(-3464,-916){\makebox(0,0)[rb]{\smash{{\SetFigFont{10}{12.0}{\rmdefault}{\mddefault}{\updefault}{\color[rgb]{0,0,0}$f_{15}, \theta_{15}$}}}}}
	\put(-2744,-61){\makebox(0,0)[b]{\smash{{\SetFigFont{10}{12.0}{\rmdefault}{\mddefault}{\updefault}{\color[rgb]{0,0,0}$b_0$}}}}}
	\put(-2744,-331){\makebox(0,0)[b]{\smash{{\SetFigFont{10}{12.0}{\rmdefault}{\mddefault}{\updefault}{\color[rgb]{0,0,0}$b_1$}}}}}
	\put(-2429,-1546){\makebox(0,0)[b]{\smash{{\SetFigFont{12}{14.4}{\rmdefault}{\mddefault}{\updefault}{\color[rgb]{0,0,0}$\times$}}}}}
	\put(-2429,-1906){\makebox(0,0)[b]{\smash{{\SetFigFont{12}{14.4}{\rmdefault}{\mddefault}{\updefault}{\color[rgb]{0,0,0}$\times$}}}}}
	\put(-2429,-2356){\makebox(0,0)[b]{\smash{{\SetFigFont{12}{14.4}{\rmdefault}{\mddefault}{\updefault}{\color[rgb]{0,0,0}$\times$}}}}}
	\put(-2834,-2176){\makebox(0,0)[rb]{\smash{{\SetFigFont{10}{12.0}{\rmdefault}{\mddefault}{\updefault}{\color[rgb]{0,0,0}$f_{15}, \omega_{15}$}}}}}
	\put(-2879,-1726){\makebox(0,0)[rb]{\smash{{\SetFigFont{10}{12.0}{\rmdefault}{\mddefault}{\updefault}{\color[rgb]{0,0,0}$f_1, \omega_1$}}}}}
	\put(-2429,-2131){\makebox(0,0)[b]{\smash{{\SetFigFont{12}{14.4}{\rmdefault}{\mddefault}{\updefault}{\color[rgb]{0,0,0}$\vdots$}}}}}
	\put(-2834,-1366){\makebox(0,0)[rb]{\smash{{\SetFigFont{10}{12.0}{\rmdefault}{\mddefault}{\updefault}{\color[rgb]{0,0,0}$f_0, \omega_0$}}}}}
	\put(-2744,659){\makebox(0,0)[b]{\smash{{\SetFigFont{10}{12.0}{\rmdefault}{\mddefault}{\updefault}{\color[rgb]{0,0,0}$a_{15}$}}}}}
	\put(-2744,-781){\makebox(0,0)[b]{\smash{{\SetFigFont{10}{12.0}{\rmdefault}{\mddefault}{\updefault}{\color[rgb]{0,0,0}$b_{15}$}}}}}
	\put(-2924,749){\makebox(0,0)[b]{\smash{{\SetFigFont{12}{14.4}{\rmdefault}{\mddefault}{\updefault}{\color[rgb]{0,0,0}$\vdots$}}}}}
	\put(-2924,-691){\makebox(0,0)[b]{\smash{{\SetFigFont{12}{14.4}{\rmdefault}{\mddefault}{\updefault}{\color[rgb]{0,0,0}$\vdots$}}}}}
	\put(-3644,-2131){\makebox(0,0)[b]{\smash{{\SetFigFont{12}{14.4}{\rmdefault}{\mddefault}{\updefault}{\color[rgb]{0,0,0}$\vdots$}}}}}
	\put(-5354,-2086){\makebox(0,0)[b]{\smash{{\SetFigFont{12}{14.4}{\rmdefault}{\mddefault}{\updefault}{\color[rgb]{0,0,0}$\ddots$}}}}}
	\put(2836,-16){\makebox(0,0)[lb]{\smash{{\SetFigFont{10}{12.0}{\rmdefault}{\mddefault}{\updefault}{\color[rgb]{0,0,0}$f$}}}}}
	\put(2836,-1456){\makebox(0,0)[lb]{\smash{{\SetFigFont{10}{12.0}{\rmdefault}{\mddefault}{\updefault}{\color[rgb]{0,0,0}$f$}}}}}
	\put(-4634,839){\makebox(0,0)[lb]{\smash{{\SetFigFont{10}{12.0}{\rmdefault}{\mddefault}{\updefault}{\color[rgb]{0,0,0}$f$}}}}}
	\put(-4634,-601){\makebox(0,0)[lb]{\smash{{\SetFigFont{10}{12.0}{\rmdefault}{\mddefault}{\updefault}{\color[rgb]{0,0,0}$f$}}}}}
	\put(-5624,-1546){\makebox(0,0)[lb]{\smash{{\SetFigFont{10}{12.0}{\rmdefault}{\mddefault}{\updefault}{\color[rgb]{0,0,0}$f$}}}}}
	\put(-5309,-1906){\makebox(0,0)[lb]{\smash{{\SetFigFont{10}{12.0}{\rmdefault}{\mddefault}{\updefault}{\color[rgb]{0,0,0}$f$}}}}}
	\put(-4859,-2446){\makebox(0,0)[lb]{\smash{{\SetFigFont{10}{12.0}{\rmdefault}{\mddefault}{\updefault}{\color[rgb]{0,0,0}$f$}}}}}
	\put(-3779,-1569){\makebox(0,0)[lb]{\smash{{\SetFigFont{10}{12.0}{\rmdefault}{\mddefault}{\updefault}{\color[rgb]{0,0,0}LPF}}}}}
	\put(-3779,-1929){\makebox(0,0)[lb]{\smash{{\SetFigFont{10}{12.0}{\rmdefault}{\mddefault}{\updefault}{\color[rgb]{0,0,0}LPF}}}}}
	\put(-3779,-2379){\makebox(0,0)[lb]{\smash{{\SetFigFont{10}{12.0}{\rmdefault}{\mddefault}{\updefault}{\color[rgb]{0,0,0}LPF}}}}}
	\put(-629,1221){\makebox(0,0)[b]{\smash{{\SetFigFont{10}{12.0}{\rmdefault}{\mddefault}{\updefault}{\color[rgb]{0,0,0}BPF}}}}}
	\put(-629,951){\makebox(0,0)[b]{\smash{{\SetFigFont{10}{12.0}{\rmdefault}{\mddefault}{\updefault}{\color[rgb]{0,0,0}BPF}}}}}
	\put(-629,501){\makebox(0,0)[b]{\smash{{\SetFigFont{10}{12.0}{\rmdefault}{\mddefault}{\updefault}{\color[rgb]{0,0,0}BPF}}}}}
	\put(-1259,1559){\makebox(0,0)[lb]{\smash{{\SetFigFont{12}{14.4}{\rmdefault}{\mddefault}{\updefault}{\color[rgb]{0,0,0}($250~\milli \kelvin$)}}}}}
	\end{picture}%
}}
\caption{
	Functional diagram of a Frequency-Domain Multiplexed (FDM) bolometer
	readout system. (For a schematic diagram, refer to
	Fig.~\ref{fig:top-level}.) 16 bolometers are multiplexed onto one set
	of wires crossing into the cryostat. High multiplexing factors are
	desirable because each wire conducts a significant amount of heat into
	the cryostat, placing a heavy burden on refrigeration. Before being
	demodulated, sky signals must be amplified by a low-noise, high-gain
	transimpedance amplifier (here, a Semiconductor Quantum Interference
	Device, or SQUID.) As SQUIDs have limited dynamic range, the signal is
	also nulled to remove residual carrier components prior to entering
	the SQUID.
	\label{fig:top-level2}
}
\end{figure*}

Figure~\ref{fig:top-level2} shows a simplified diagram of a FDM bolometer
readout system. The goal of such a system is to minimize the number of wires
crossing into the cryostat by multiplexing several bolometer signals on a
single set of wires, without degrading each bolometer's noise performance.
Multiplexing is desirable since the thermal load presented by wires entering
the cryostat is substantial. Cryogenic refrigeration consumes a significant
fraction of an experiment's power budget, and cryogenic fridge cycling is
time-consuming enough to impact the experiment's uptime.

This system synthesizes a set of carrier combs, which enter the cryostat
on a single wire. Each of these combs is selected by a corresponding bandpass
filter and passes through a bolometer, which imposes a gain proportional to
its incoming photon flux. This time-varying gain modulates each carrier
frequency, producing sidebands. Each bolometer's band-limited output is summed
together to form the modulated sky signal.

The modulated sky signal is of very low power, and must be amplified
before being passed to room-temperature electronics. To do so, we use a
Semiconductor Quantum Interference Device (SQUID). SQUIDs may be used as
extremely sensitive transimpedance amplifiers, but are only approximately
linear in a limited dynamic range. To minimize harmonic distortion, we operate
the SQUID in a negative feedback loop. In addition, we remove the residual
carrier signals via destructive interference using a nuller comb.

Finally, the nulled sky signal is passed back to room-temperature
electronics for demodulation. Each channel is demodulated to baseband,
low-pass filtered, and archived for analysis.

Our design has much in common with the digital up- and
down-converters (DUC and DDC, respectively) used in GSM and other multichannel
base-station applications~\cite{Andraka2000, Ogilvie2006}. However, the DFMUX
differs from traditional DUC and DDC implementations in several crucial ways:
\begin{itemize}
\item Our bandwidth for each channel is relatively small (10s of Hz).
  Since the modulator sinusoid is used to both electrically
  bias the TES and carry the ultra-low noise signals, the dynamic
  range requirement of $\sim 10^6 \sqrt{\mathrm{Hz}}$ is extreme.
  We require no degradation of noise performance down to several 10s
  of mHz.
\item The modulator generates only combs of carrier sinusoids. The modulation
process itself takes place inside the cryogenic vessel.
\item Both the modulator and demodulator for each channel reside \emph{on the
same} FPGA. This means, for example, that each channel's modulator and
demodulator need only be synchronized once since their clocks cannot drift
with respect to each other. Clock jitter cancels and is not a
significant source of low-frequency noise.
\item In addition to its readout and tuning roles, the DFMUX is also used as a
real-time bench instrument. Thus, it is desirable to have flexible control and
a readily understandable signal path. This motivates features that are used
occasionally, such as input re-routing and debugging facilities.
\item The system must be robust against crashes causing interruptions
  in the detector biases since this can necessitate thermal cycling of
  the sensors, requiring re-initialization of cryogenic systems.
This can take about an hour per comb, adversely affecting the
observing efficiency of the telescope applications.
\end{itemize}

\section{Signal Path
	\label{sec:signalpath}
}

\begin{figure}
\centerline{\scalebox{0.65}{
		\begin{picture}(0,0)%
	\includegraphics{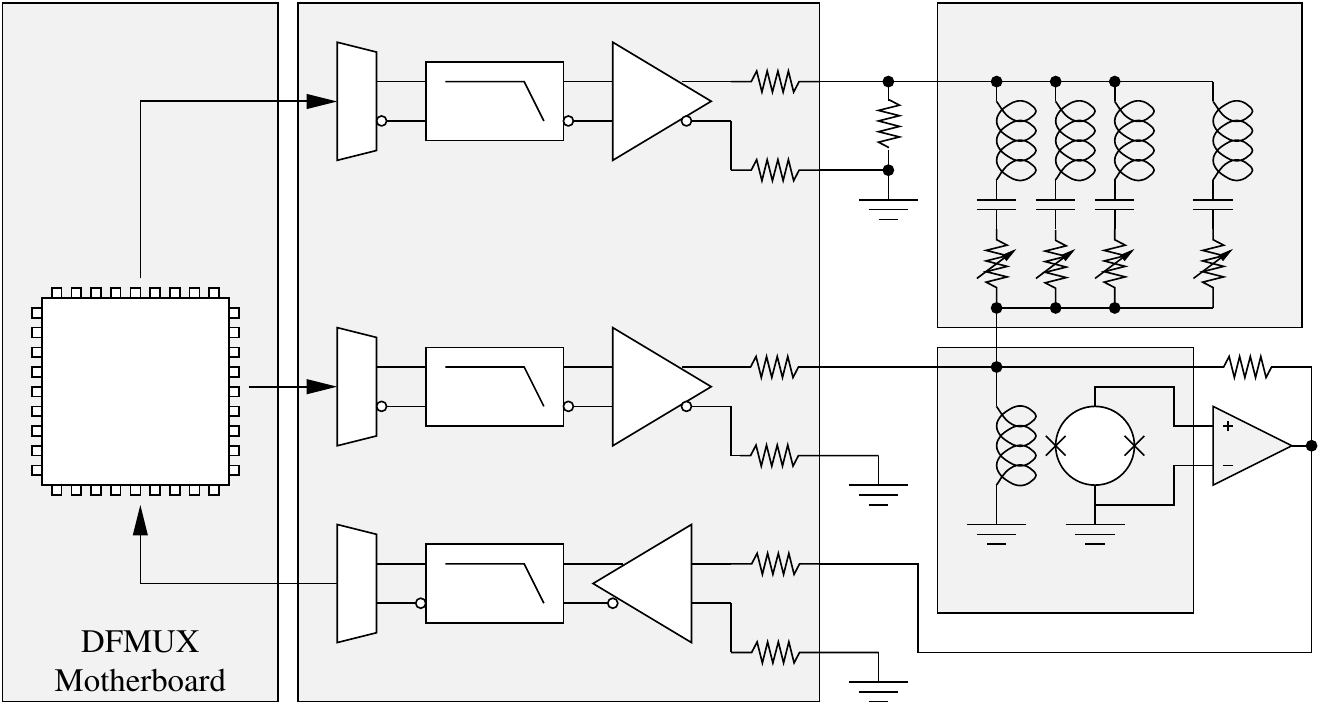}%
	\end{picture}%
	\setlength{\unitlength}{4144sp}%
	\begingroup\makeatletter\ifx\SetFigFont\undefined%
	\gdef\SetFigFont#1#2#3#4#5{%
	  \normalfont\fontsize{#1}{#2pt}%
	  \fontfamily{#3}\fontseries{#4}\fontshape{#5}%
	  \selectfont}%
	\fi\endgroup%
	\begin{picture}(6028,3222)(169,-1876)
	\put(4051,-466){\makebox(0,0)[b]{\smash{{\SetFigFont{8}{9.6}{\rmdefault}{\mddefault}{\updefault}{\color[rgb]{0,0,0}$+$}}}}}
	\put(4051,-590){\makebox(0,0)[b]{\smash{{\SetFigFont{8}{9.6}{\rmdefault}{\mddefault}{\updefault}{\color[rgb]{0,0,0}$V_N$}}}}}
	\put(4051,-691){\makebox(0,0)[b]{\smash{{\SetFigFont{8}{9.6}{\rmdefault}{\mddefault}{\updefault}{\color[rgb]{0,0,0}$-$}}}}}
	\put(4051,-1366){\makebox(0,0)[b]{\smash{{\SetFigFont{8}{9.6}{\rmdefault}{\mddefault}{\updefault}{\color[rgb]{0,0,0}$+$}}}}}
	\put(4051,-1490){\makebox(0,0)[b]{\smash{{\SetFigFont{8}{9.6}{\rmdefault}{\mddefault}{\updefault}{\color[rgb]{0,0,0}$V_D$}}}}}
	\put(4051,-1591){\makebox(0,0)[b]{\smash{{\SetFigFont{8}{9.6}{\rmdefault}{\mddefault}{\updefault}{\color[rgb]{0,0,0}$-$}}}}}
	\put(4051,839){\makebox(0,0)[b]{\smash{{\SetFigFont{8}{9.6}{\rmdefault}{\mddefault}{\updefault}{\color[rgb]{0,0,0}$+$}}}}}
	\put(4051,715){\makebox(0,0)[b]{\smash{{\SetFigFont{8}{9.6}{\rmdefault}{\mddefault}{\updefault}{\color[rgb]{0,0,0}$V_C$}}}}}
	\put(4051,614){\makebox(0,0)[b]{\smash{{\SetFigFont{8}{9.6}{\rmdefault}{\mddefault}{\updefault}{\color[rgb]{0,0,0}$-$}}}}}
	\put(4321,749){\rotatebox{270.0}{\makebox(0,0)[b]{\smash{{\SetFigFont{8}{9.6}{\rmdefault}{\mddefault}{\updefault}{\color[rgb]{0,0,0}$R_\mathrm{bias}$}}}}}}
	\put(1801,434){\makebox(0,0)[b]{\smash{{\SetFigFont{10}{12.0}{\rmdefault}{\mddefault}{\updefault}{\color[rgb]{0,0,0}Carrier}}}}}
	\put(1801,-61){\makebox(0,0)[b]{\smash{{\SetFigFont{10}{12.0}{\rmdefault}{\mddefault}{\updefault}{\color[rgb]{0,0,0}Nuller}}}}}
	\put(1801,-961){\makebox(0,0)[b]{\smash{{\SetFigFont{10}{12.0}{\rmdefault}{\mddefault}{\updefault}{\color[rgb]{0,0,0}Demod}}}}}
	\put(4501,479){\rotatebox{270.0}{\makebox(0,0)[b]{\smash{{\SetFigFont{10}{12.0}{\rmdefault}{\mddefault}{\updefault}{\color[rgb]{0,0,0}Bolometer 1}}}}}}
	\put(5986,479){\rotatebox{270.0}{\makebox(0,0)[b]{\smash{{\SetFigFont{10}{12.0}{\rmdefault}{\mddefault}{\updefault}{\color[rgb]{0,0,0}Bolometer 16}}}}}}
	\put(788,-556){\makebox(0,0)[b]{\smash{{\SetFigFont{10}{12.0}{\rmdefault}{\mddefault}{\updefault}{\color[rgb]{0,0,0}FPGA}}}}}
	\put(788,-376){\makebox(0,0)[b]{\smash{{\SetFigFont{10}{12.0}{\rmdefault}{\mddefault}{\updefault}{\color[rgb]{0,0,0}Virtex-4}}}}}
	\put(1756,-421){\rotatebox{270.0}{\makebox(0,0)[b]{\smash{{\SetFigFont{10}{12.0}{\rmdefault}{\mddefault}{\updefault}{\color[rgb]{0,0,0}DAC}}}}}}
	\put(3151,-1366){\makebox(0,0)[b]{\smash{{\SetFigFont{12}{14.4}{\rmdefault}{\mddefault}{\updefault}{\color[rgb]{0,0,0}$G_D$}}}}}
	\put(3151,839){\makebox(0,0)[b]{\smash{{\SetFigFont{12}{14.4}{\rmdefault}{\mddefault}{\updefault}{\color[rgb]{0,0,0}$G_C$}}}}}
	\put(3151,-466){\makebox(0,0)[b]{\smash{{\SetFigFont{12}{14.4}{\rmdefault}{\mddefault}{\updefault}{\color[rgb]{0,0,0}$G_N$}}}}}
	\put(2701,-1816){\makebox(0,0)[b]{\smash{{\SetFigFont{10}{12.0}{\rmdefault}{\mddefault}{\updefault}{\color[rgb]{0,0,0}Mezzanine Card ($\times 2$)}}}}}
	\put(1756,884){\rotatebox{270.0}{\makebox(0,0)[b]{\smash{{\SetFigFont{10}{12.0}{\rmdefault}{\mddefault}{\updefault}{\color[rgb]{0,0,0}DAC}}}}}}
	\put(2251,839){\makebox(0,0)[lb]{\smash{{\SetFigFont{10}{12.0}{\rmdefault}{\mddefault}{\updefault}{\color[rgb]{0,0,0}LPF}}}}}
	\put(2251,-466){\makebox(0,0)[lb]{\smash{{\SetFigFont{10}{12.0}{\rmdefault}{\mddefault}{\updefault}{\color[rgb]{0,0,0}LPF}}}}}
	\put(1756,-1321){\rotatebox{270.0}{\makebox(0,0)[b]{\smash{{\SetFigFont{10}{12.0}{\rmdefault}{\mddefault}{\updefault}{\color[rgb]{0,0,0}ADC}}}}}}
	\put(2251,-1366){\makebox(0,0)[lb]{\smash{{\SetFigFont{10}{12.0}{\rmdefault}{\mddefault}{\updefault}{\color[rgb]{0,0,0}LPF}}}}}
	\put(5581,659){\makebox(0,0)[b]{\smash{{\SetFigFont{10}{12.0}{\rmdefault}{\mddefault}{\updefault}{\color[rgb]{0,0,0}$\cdots$}}}}}
	\put(5491, 74){\makebox(0,0)[b]{\smash{{\SetFigFont{10}{12.0}{\rmdefault}{\mddefault}{\updefault}{\color[rgb]{0,0,0}$\cdots$}}}}}
	\put(5311,1109){\makebox(0,0)[b]{\smash{{\SetFigFont{10}{12.0}{\rmdefault}{\mddefault}{\updefault}{\color[rgb]{0,0,0}Sub-Kelvin Stage}}}}}
	\put(5041,-1321){\makebox(0,0)[b]{\smash{{\SetFigFont{10}{12.0}{\rmdefault}{\mddefault}{\updefault}{\color[rgb]{0,0,0}4\degree K Stage}}}}}
	\end{picture}%
	}}
\caption{
	Simplified schematic of one readout module multiplexing 16 TES
	detectors on a single set of cryogenic wires.
	\label{fig:top-level}
}
\end{figure}
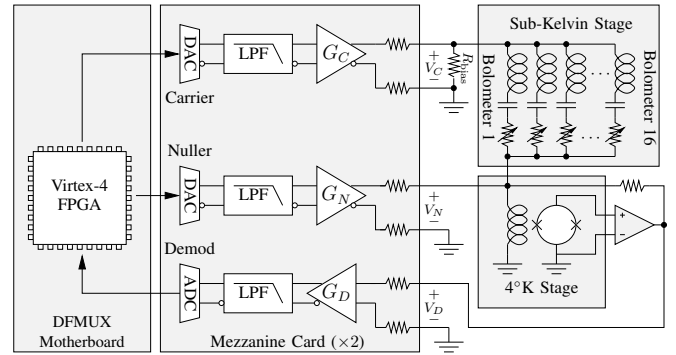

In this section, we describe the signal path for each detector channel during
ordinary operation.  Fig.~\ref{fig:top-level} shows a schematic including a
single readout module.\footnote{
	Fig.~\ref{fig:top-level} has been simplified. It neglects, for
	example, details of the differential signaling scheme and a separate
	board containing additional amplifiers (the ``SQUID controller''
	board.) For details of the warm and cold electronics beyond the DFMUX,
	readers are referred to~\cite{Aubin2010, Reichborn2010}.
}
At a channel's resonant frequency, its LC filter has negligible
impedance, permitting current to flow through the associated bolometer
(modeled here as a resistance that varies with incoming optical power.)
Although crosstalk from adjacent channels is nonzero, it may be controlled by
selecting a set of resonant frequencies with adequate spacing.

Each DFMUX consists of a motherboard, shown on the left side of
Fig.~\ref{fig:top-level}, and up to two mezzanines. Each mezzanine contains
the analog electronics for two of the DFMUX's four modules, including 16-bit
Digital-to-Analog Converters (DACs) and 14-bit Analog-to-Digital Converters
(ADCs) operating at 25 MSPS. Each mezzanine also houses analog anti-aliasing
filters at both DAC outputs and ADC inputs, as well as a sequence of
amplifiers with programmable gains.  The room-temperature backend electronics
presently supports up to 16 multiplexed TESes per module.  Instruments that
are presently operating have just 8~TESes wired together per module; upgrades
to a larger channel count are in progress.

The conversion between ADC and DAC units and physical quantities are
summarized in Table~\ref{tab:mezz-conversions}. Voltages are measured at the
mezzanine connectors (i.e. $V_C$, $V_N$, and $V_D$ shown in
Fig.~\ref{fig:top-level}), and assuming matched load resistances. The
post-ADC signal chain in the DMFD involves other conversion factors that are
examined below; these factors are not present in
Table~\ref{tab:mezz-conversions}.

\begin{table}
\centerline{\begin{tabular}{l|c|c|c|r} \hline
& \multicolumn{2}{|c|}{\bf{DMFS}} & \multicolumn{2}{|c}{\bf{DMFD}} \\ \hline
\bf{Setting}	& \bf{PGA} & \bf{DAC Scale} & \bf{PGA} & \bf{ADC Scale} \\
		& \bf{Gain} & ($\volt/\mathrm{LSB}$) & \bf{Gain} & ($\mathrm{LSB}/\volt$) \\ \hline
Low		& 0.48	& $3.63 \times 10^{-6}$	& 0.99	& $16,222$ \\
Med. Low	& 1.09	& $8.29 \times 10^{-6}$	& 4.76	& $78,019$ \\
Med. High	& 3.33	& $2.54 \times 10^{-5}$	& 33.3	& $546,133$ \\
High		& 10	& $7.63 \times 10^{-5}$	& 100	& $1,638,400$ \\ \hline
\end{tabular}}

% Calculations for DMFS:
%
% 0.5v/2**16	(DAC Vpp / 16 bits)
%	* 2	(fixed gain)
%	* [ 1000./2100 1000./920 1000./300 1000./100 ]	(PGA)
%	* 0.5	(assume 100-ohm load, measure at mezz out)
% = [	3.63304501e-06,
%	8.29282014e-06,
%	2.54313151e-05,
%	7.62939453e-05 ]
%
% Calculations for DMFD:
%
% 2		(fixed gain)
%	* [ 10000./10100 10000./2100 10000./300 10000./100 ]	(PGA)
%	* 0.5	(Chebyshev filter)
%	* 2**14 / 2	(ADC 14 bits / 2 Vpp)
% = [	16221.782178217822
%	78019.047619047618
%	546133.33333333337
%	1638400.0

\vspace{0.5em}
\caption{
	Mezzanine conversion factors. The gains for the carrier, nuller, and
	demodulator amplifiers are independently controlled in software.
	Voltages are specified at the mezzanine's output (carrier/nuller) or
	input (demodulator). ADC and DAC counts are specified at the
	converter itself; additional conversions due to the signal chain are
	described below.
	\label{tab:mezz-conversions}
}
\end{table}

The carrier DAC generates a weighted sum of sinusoids. Each sinusoidal
component is tuned to match the resonant frequency of an LC filter in the
cryostat associated with a single bolometer, with an amplitude chosen to
provide the necessary voltage bias for the TES.  The tuning procedures used to
optimize the bias voltages and currents for the detectors and SQUIDs are
described in~\cite{Macdermid2009}.  The resistance of each bolometer (and thus
the current through it) varies in response to sky signals, amplitude
modulating the sinusoidal carrier. As a result, each bolometer adds sidebands
around its carrier frequency with a typical signal bandwidth of
$0.01-100~\hertz$.

Currents from all bolometers are summed at the input inductor of the SQUID. A
second (nuller) DAC generates a second weighted sum of sinusoids that is
injected at the SQUID input. The phase and amplitude of these sinusoids are
adjusted to cancel the carrier signals in order to reduce the dynamic range
requirements of the SQUID.  Were nulling perfect, the current through the
SQUID-coupled input inductor would consist only of the sidebands from each
bolometer. In reality, some amount of carrier power remains. The SQUID is
operated in a flux-locked loop with shunt feedback~\cite{Spieler02}. It
converts current through its input inductor to a voltage signal at its output.
This signal is amplified, filtered to limit aliasing during sampling,
digitized by the ADC, and sent to the FPGA for demodulation.

In the following sections, we describe the structures within the FPGA that
perform each of the biasing and readout functions. We refer to the signal path
performing carrier synthesis as the Digital Multi-Frequency Synthesizer
(DMFS) and the demodulation signal path as the Digital Multi-Frequency
Demodulator (DMFD).

\subsection{Digital Multi-Frequency Synthesizer (DMFS)}

\subsubsection{Structure}

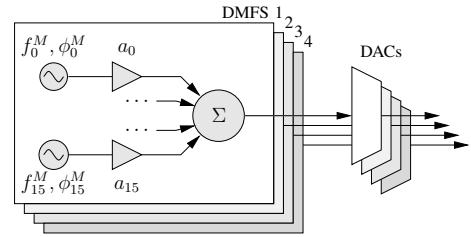
\begin{figure}
\centerline{\scalebox{0.65}{
	\begin{picture}(0,0)%
	\includegraphics{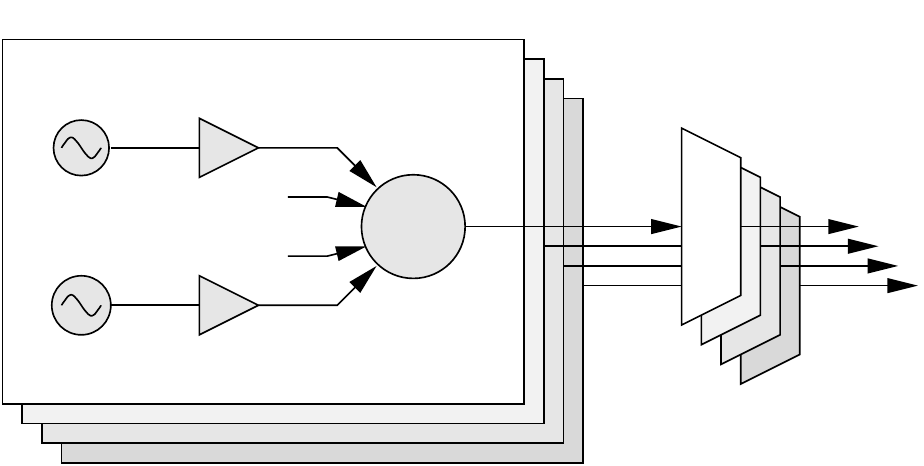}%
	\end{picture}%
	\setlength{\unitlength}{4144sp}%
	\begingroup\makeatletter\ifx\SetFigFont\undefined%
	\gdef\SetFigFont#1#2#3#4#5{%
	  \reset@font\fontsize{#1}{#2pt}%
	  \fontfamily{#3}\fontseries{#4}\fontshape{#5}%
	  \selectfont}%
	\fi\endgroup%
	\begin{picture}(4209,2112)(169,-973)
	\put(2746,929){\makebox(0,0)[rb]{\smash{{\SetFigFont{10}{12.0}{\rmdefault}{\mddefault}{\updefault}{\color[rgb]{0,0,0}2}}}}}
	\put(2836,839){\makebox(0,0)[rb]{\smash{{\SetFigFont{10}{12.0}{\rmdefault}{\mddefault}{\updefault}{\color[rgb]{0,0,0}3}}}}}
	\put(2926,749){\makebox(0,0)[rb]{\smash{{\SetFigFont{10}{12.0}{\rmdefault}{\mddefault}{\updefault}{\color[rgb]{0,0,0}4}}}}}
	\put(2656,1019){\makebox(0,0)[rb]{\smash{{\SetFigFont{10}{12.0}{\rmdefault}{\mddefault}{\updefault}{\color[rgb]{0,0,0}DMFS 1}}}}}
	\put(3556,659){\makebox(0,0)[b]{\smash{{\SetFigFont{10}{12.0}{\rmdefault}{\mddefault}{\updefault}{\color[rgb]{0,0,0}DACs}}}}}
	\put(1306,209){\makebox(0,0)[b]{\smash{{\SetFigFont{12}{14.4}{\rmdefault}{\mddefault}{\updefault}{\color[rgb]{0,0,0}$\cdots$}}}}}
	\put(1306,-61){\makebox(0,0)[b]{\smash{{\SetFigFont{12}{14.4}{\rmdefault}{\mddefault}{\updefault}{\color[rgb]{0,0,0}$\cdots$}}}}}
	\put(1216,704){\makebox(0,0)[b]{\smash{{\SetFigFont{12}{14.4}{\rmdefault}{\mddefault}{\updefault}{\color[rgb]{0,0,0}$a_0$}}}}}
	\put(541,-556){\makebox(0,0)[b]{\smash{{\SetFigFont{12}{14.4}{\rmdefault}{\mddefault}{\updefault}{\color[rgb]{0,0,0}$f^M_{15}, \phi^M_{15}$}}}}}
	\put(1216,-556){\makebox(0,0)[b]{\smash{{\SetFigFont{12}{14.4}{\rmdefault}{\mddefault}{\updefault}{\color[rgb]{0,0,0}$a_{15}$}}}}}
	\put(2062, 60){\makebox(0,0)[b]{\smash{{\SetFigFont{12}{14.4}{\rmdefault}{\mddefault}{\updefault}{\color[rgb]{0,0,0}$\Sigma$}}}}}
	\put(541,704){\makebox(0,0)[b]{\smash{{\SetFigFont{12}{14.4}{\rmdefault}{\mddefault}{\updefault}{\color[rgb]{0,0,0}$f^M_0, \phi^M_0$}}}}}
	\end{picture}%
	}}%\input{schematic_dmfs.tex}}}
\caption{
	Carriers are generated by four identical DMFS blocks, each of which
	synthesizes a carrier comb for one DAC. An identical system is used
	for nuller synthesis.
	\label{fig:dmfs}
}
\end{figure}

Each of the four modules in a DFMUX is associated with two identical DMFS
blocks. One generates the carrier signals that bias bolometers. The other
DMFS generates the nuller signals that cancel out these carriers at the SQUID
input.  These blocks synthesize waveforms for one wire at 25 MSPS. DMFSes
operate independently and are not synchronized (meaning that the fixed phase
between any two DMFS channels set to the same frequency cannot be determined
without measurement.) The schematic for a single DMFS block is shown in
Fig.~\ref{fig:dmfs}.

Each of the 16 sinusoids in an individual DMFS block are generated using
vendor-supplied~\cite{XilinxDDS2007}, 12-bit, 2's-complement Direct Digital
Synthesizers (DDSes) as shown in Fig.~\ref{fig:dmfs}. To maximize design
density, each DDS generates sinusoids for 8 channels and operates at an
internal clock rate of 200 MHz.  Each DDS uses a 32-bit phase accumulator to
track the waveform's phase.  Each clock cycle this accumulator is incremented,
truncated to 14 bits, then used to reference the address of the waveform's
amplitude in a look-up table. A separate 32 bit register provides a
programmable phase offset. When operating at 25 MHz, the frequency can be
specified to 0.006 Hz. The spurious free dynamic range (SFDR) is 96 dB.
This algorithm does not use substantial logic resources, but does use a
significant number of the FPGA's block RAM (BRAM) structures. Since each DDS
block synthesizes 8 sinusoids, two such DDSes are required for each 16-channel
module.

After synthesis by a DDS, each channel's sinusoid is then weighted by a 20-bit
2's-complement amplitude. This weighting is used to allow Joule heating
provided by the voltage bias to be individually controlled. This allows each
bolometer, which can have significantly different device parameters due to
fabrication process variability, to be tuned to its optimum bias point. Each
weighted sinusoid is then summed with the other 15 channels in its DMFS, and
truncated to form a 16-bit signal for the associated DAC.  The signal from
each DMFS is finally converted from 2's-complement to a positive binary number
and passed to the 25 MHz clock domain for digital-to-analog conversion.

The slight DC bias resulting from truncation in the DMFS is unimportant
since carrier and nuller outputs are AC-coupled.

\subsection{Digital Multi-Frequency Demodulator (DMFD)}

\begin{figure*}
\centerline{\scalebox{0.75}{
	\begin{picture}(0,0)%
	\includegraphics{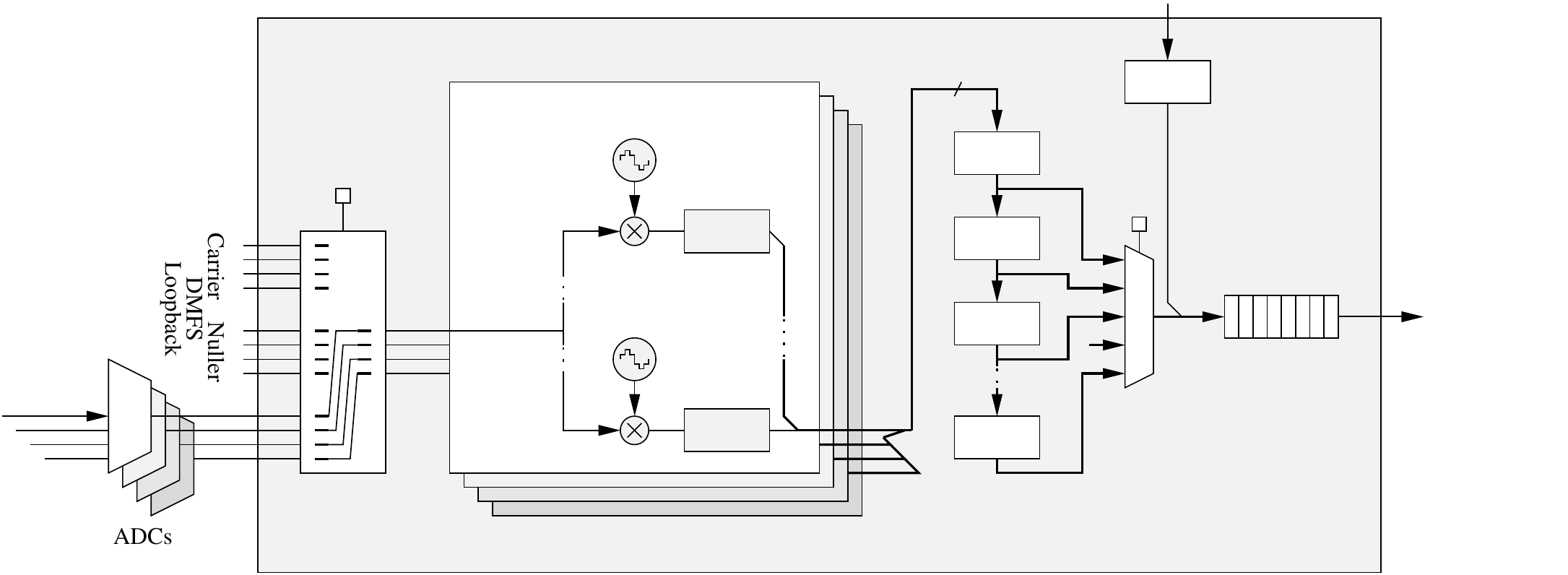}%
	\end{picture}%
	\setlength{\unitlength}{4144sp}%
	\begingroup\makeatletter\ifx\SetFigFont\undefined%
	\gdef\SetFigFont#1#2#3#4#5{%
	  \reset@font\fontsize{#1}{#2pt}%
	  \fontfamily{#3}\fontseries{#4}\fontshape{#5}%
	  \selectfont}%
	\fi\endgroup%
	\begin{picture}(9914,3624)(-4511,-703)
	\put(-2339,1874){\makebox(0,0)[b]{\smash{{\SetFigFont{12}{14.4}{\rmdefault}{\mddefault}{\updefault}{\color[rgb]{0,0,0}Crossbar}}}}}
	\put(-2339,2054){\makebox(0,0)[b]{\smash{{\SetFigFont{12}{14.4}{\rmdefault}{\mddefault}{\updefault}{\color[rgb]{0,0,0}Routing}}}}}
	\put(811,2369){\makebox(0,0)[rb]{\smash{{\SetFigFont{12}{14.4}{\rmdefault}{\mddefault}{\updefault}{\color[rgb]{0,0,0}2}}}}}
	\put(901,2279){\makebox(0,0)[rb]{\smash{{\SetFigFont{12}{14.4}{\rmdefault}{\mddefault}{\updefault}{\color[rgb]{0,0,0}3}}}}}
	\put(991,2189){\makebox(0,0)[rb]{\smash{{\SetFigFont{12}{14.4}{\rmdefault}{\mddefault}{\updefault}{\color[rgb]{0,0,0}4}}}}}
	\put(3601,1199){\makebox(0,0)[b]{\smash{{\SetFigFont{12}{14.4}{\rmdefault}{\mddefault}{\updefault}{\color[rgb]{0,0,0}FIFO}}}}}
	\put(3646,2459){\makebox(0,0)[b]{\smash{{\SetFigFont{12}{14.4}{\rmdefault}{\mddefault}{\updefault}{\color[rgb]{0,0,0}Timestamp}}}}}
	\put(3646,2279){\makebox(0,0)[b]{\smash{{\SetFigFont{12}{14.4}{\rmdefault}{\mddefault}{\updefault}{\color[rgb]{0,0,0}Register}}}}}
	\put(721,2459){\makebox(0,0)[rb]{\smash{{\SetFigFont{12}{14.4}{\rmdefault}{\mddefault}{\updefault}{\color[rgb]{0,0,0}Mux 1}}}}}
	\put(4996,884){\makebox(0,0)[b]{\smash{{\SetFigFont{12}{14.4}{\rmdefault}{\mddefault}{\updefault}{\color[rgb]{0,0,0}Processor}}}}}
	\put(4996,659){\makebox(0,0)[b]{\smash{{\SetFigFont{12}{14.4}{\rmdefault}{\mddefault}{\updefault}{\color[rgb]{0,0,0}Bus}}}}}
	\put(4996,1109){\makebox(0,0)[b]{\smash{{\SetFigFont{12}{14.4}{\rmdefault}{\mddefault}{\updefault}{\color[rgb]{0,0,0}To}}}}}
	\put(1531,2459){\makebox(0,0)[b]{\smash{{\SetFigFont{10}{12.0}{\rmdefault}{\mddefault}{\updefault}{\color[rgb]{0,0,0}64 channels}}}}}
	\put( 91,1424){\makebox(0,0)[b]{\smash{{\SetFigFont{12}{14.4}{\rmdefault}{\mddefault}{\updefault}{\color[rgb]{0,0,0}CIC1}}}}}
	\put( 91,164){\makebox(0,0)[b]{\smash{{\SetFigFont{12}{14.4}{\rmdefault}{\mddefault}{\updefault}{\color[rgb]{0,0,0}CIC1}}}}}
	\put(-494,929){\makebox(0,0)[b]{\smash{{\SetFigFont{12}{14.4}{\rmdefault}{\mddefault}{\updefault}{\color[rgb]{0,0,0}$f^D_{16}, \phi^D_{16}$}}}}}
	\put(-494,2189){\makebox(0,0)[b]{\smash{{\SetFigFont{12}{14.4}{\rmdefault}{\mddefault}{\updefault}{\color[rgb]{0,0,0}$f^D_0, \phi^D_0$}}}}}
	\put(-494,1109){\makebox(0,0)[b]{\smash{{\SetFigFont{12}{14.4}{\rmdefault}{\mddefault}{\updefault}{\color[rgb]{0,0,0}$\vdots$}}}}}
	\put(1801,1919){\makebox(0,0)[b]{\smash{{\SetFigFont{12}{14.4}{\rmdefault}{\mddefault}{\updefault}{\color[rgb]{0,0,0}CIC2}}}}}
	\put(1801,1379){\makebox(0,0)[b]{\smash{{\SetFigFont{12}{14.4}{\rmdefault}{\mddefault}{\updefault}{\color[rgb]{0,0,0}FIR1}}}}}
	\put(1801,839){\makebox(0,0)[b]{\smash{{\SetFigFont{12}{14.4}{\rmdefault}{\mddefault}{\updefault}{\color[rgb]{0,0,0}FIR2}}}}}
	\put(1801,119){\makebox(0,0)[b]{\smash{{\SetFigFont{12}{14.4}{\rmdefault}{\mddefault}{\updefault}{\color[rgb]{0,0,0}FIR6}}}}}
	\put(-2924,1154){\makebox(0,0)[rb]{\smash{{\SetFigFont{20}{24.0}{\rmdefault}{\mddefault}{\updefault}{\color[rgb]{0,0,0}\{}}}}}
	\put(-2924,614){\makebox(0,0)[rb]{\smash{{\SetFigFont{20}{24.0}{\rmdefault}{\mddefault}{\updefault}{\color[rgb]{0,0,0}\{}}}}}
	\end{picture}%
}}
\caption{
	Simplified schematic showing the Digital Multi-Frequency Demodulator
	(DMFD) signal path. One DMFD services 4 readout modules.
	\label{fig:dmfd}
}
\end{figure*}
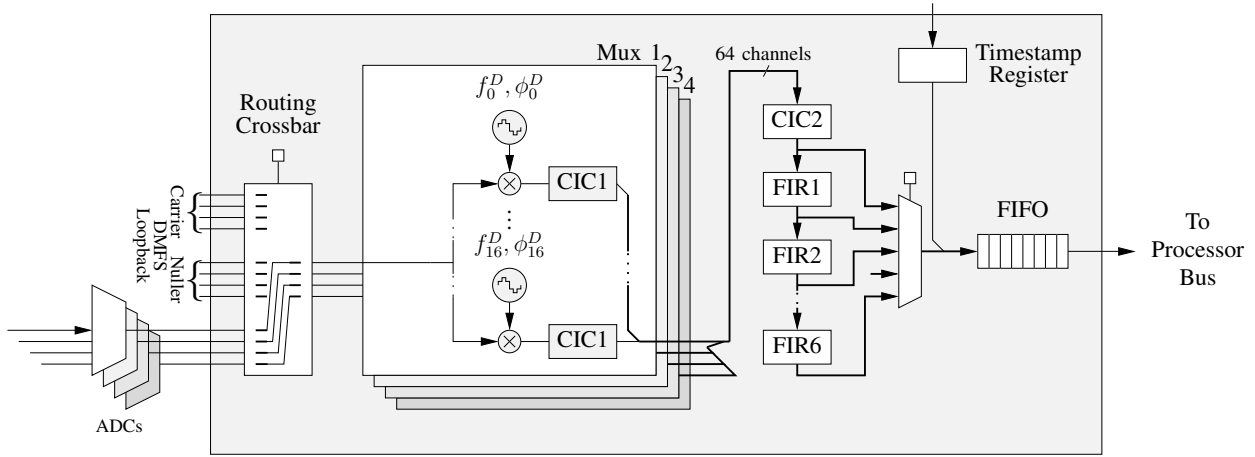

Signal processing within the DMFD is illustrated in Fig.~\ref{fig:dmfd}.

\subsubsection{Routing}

The DMFD simultaneously demodulates input signals from four readout modules,
each sampled at 25 MSPS.  These inputs generally correspond to each of the
DFMUX's four 14-bit ADCs.  For debugging and biasing duties, it also is
possible to reroute the four inputs directly to the DMFS signals, bypassing
A/D and D/A conversion entirely. (This routing capability has been used, for
example, to use the DFMUX as a network-analysis tool, comparing the phase
shift and attenuation caused by an analog circuit to a reference sinusoid that
has been rerouted from the DMFS directly into a separate DMFD.)

\subsubsection{Downconversion}

Once each signal has been routed for demodulation, it is mixed with 17
real reference waveforms to produce baseband signals. The frequency and phase
of the reference waveforms can be programmed independently. The number of
demodulators is one greater than the number of multiplexed detector channels.
The 17th channel is identical to the other 16, but is not otherwise committed
to a bolometer and is thus free to be placed in quadrature with another
channel's demodulator to form a complex (IQ) demodulator. A complex
demodulator is helpful during tuning and testing, but is not necessary during
ordinary readout.

Waveforms generated within the DMFD are equivalent to coarsely quantized
sinusoids. These waveforms are synthesized with a custom DDS that uses samples
of a length-16 sequence addressed with a phase accumulator at the desired
frequency. The frequency spectra of references generated in this manner are
tractable but complicated~\cite{Jenq1988a, Jenq1988}, and are suitably
approximated by that of ideal sinusoids for our purposes. The sequence is
symmetric; a positive half-cycle is $[0,3,6,7,7,7,6,3]/8$. The RMS amplitude
of waveforms synthesized using this sequence is approximately $0.681$, or
about 4\% lower than the ideal sinusoid's RMS of $1/\sqrt{2}$. The
synthesizer's SNR is frequency-dependent.  Vendor-supplied DDSes are not used
for downconversion in the DMFD due to limited FPGA resources, and because the
available DDSes do not permit phase synchronization between channels.

In order to permit quadrature demodulation, waveforms synthesized in the DMFD may
be phase locked. Each synthesizer's instantaneous phase may be placed on a
bus, and other synthesizers may load it if instructed. This process allows any
two channels in the DMFD to be phase-locked with a fixed, programmable phase
offset between them. It is typically used to construct quadrature demodulators
using the 17th DMFD channel mentioned above. While this implementation does
require that all channels share a phase bus, it does so at modest cost in
terms of FPGA resources, and uses vastly fewer resources than producing
quadrature outputs for each channel.

\subsubsection{Decimation}

After mixing, each channel within the DMFD contains a demodulated signal at
baseband, sampled at 25 MSPS. Since the bandwidth of interest is a tiny
fraction of the full Nyquist bandwidth, and since this sampling rate presents
practical difficulties for streaming and storage, it is desirable to decimate
these signals by a large factor ($\sim 10^5$). The remainder of the DMFD is
largely devoted to performing this decimation efficiently.

Each channel is first processed by its own Cascaded Integrator-Comb (CIC)
decimation filter~\cite{Hogenauer1981}, labelled CIC1 in Fig.~\ref{fig:dmfd}.
CIC decimators are constructed using only accumulators and adders, which map
very naturally to the resources available on an FPGA. However, since CICs
exhibit significant passband nonlinarity and have broad transition regions,
they are generally followed by compensating FIRs to restore uniform gain and
perform additional decimation. The first-stage CIC reduces the overall
sampling rate by a factor of 128, has 3 stages, and uses 35 bits internally.
CIC1 output is then truncated to 17 bits.

After the CIC1 filter, all 68 channels are multiplexed onto a shared data
bus.  Hereafter, all filters are time-multiplexed and operate on each channel
in sequence for a substantial savings in FPGA resources.

Following this multiplexer, a second CIC filter (CIC2) decimates by a further
factor of 16. CIC2 uses four stages using 33-bit internal signals. After CIC2,
each channel has a data rate of $12.21~\kilo\hertz$ and is truncated to 17
bits.

\begin{table}
\centerline{\begin{tabular}{c|c|c|c|c} \hline
\bf{Filter} & \bf{Decimation} & \bf{Taps} & \bf{Output} & \bf{Signal} \\
& \bf{Factor} & & \bf{Rate} & \bf{Bandwidth} \\ \hline
CIC1 & 128 & - & 195.3~\kilo \hertz & $1.37~\kilo \hertz$ \\
CIC2 & 16 & - & 12.21~\kilo \hertz & $1.37~\kilo \hertz$ \\
FIR1 & 2 & 43 & 6103~\hertz & $1.37~\kilo \hertz$ \\
FIR2 & 2 & 108 & 3051~\hertz & $1.22~\kilo \hertz$ \\
FIR3 & 2 & 108 & 1526~\hertz & $610~\hertz$ \\
FIR4 & 2 & 108 & 762.9~\hertz & $305~\hertz$ \\
FIR5 & 2 & 108 & 381.5~\hertz & $153~\hertz$ \\
FIR6 & 2 & 108 & 190.7~\hertz & $76~\hertz$ \\ \hline
\end{tabular}}
\vspace{0.5em}
\caption{
	Decimation stages in the demodulator chain. The demodulator output may
	be selected from CIC2 or any of the FIRs, and includes timestamps.
	``Bandwidth'' indicates the useful bandwidth of the stage, which is
	guaranteed to be free of spurious components to about $-100~\deci
	\bel$.
	\label{tab:decimation-stages}
}
\end{table}

The signals are then processed by a sequence of Finite Impulse Response (FIR)
filters~\cite{XilinxFIR2007}. The first FIR includes correction for amplitude distortion introduced
by the CIC filtering stages, and is 43 taps in length. The second through
sixth FIRs are identical, unit-gain, 108-tap filters with unit passband gain.
Although it would be computationally more efficient to optimize each FIR
individually, using the same FIR prototype for each of the DMFD's possible
outputs provides a consistent and predictable output scale at a variety of
sampling rates.  This consistency is particularly desirable since switching
between FIRs is only typically used during debugging, when it is sometimes
necessary to view the baseband signal with greater bandwidth than is provided
by normal readout (which typically occurs at FIR5 or FIR6.) The DMFD's
decimation stages are summarized in Table~\ref{tab:decimation-stages}. Each
filter's accumulator width is large enough to avoid overflow. Output is
truncated to 17 bits.

\begin{figure*}
\centerline{\includegraphics[width=\textwidth]{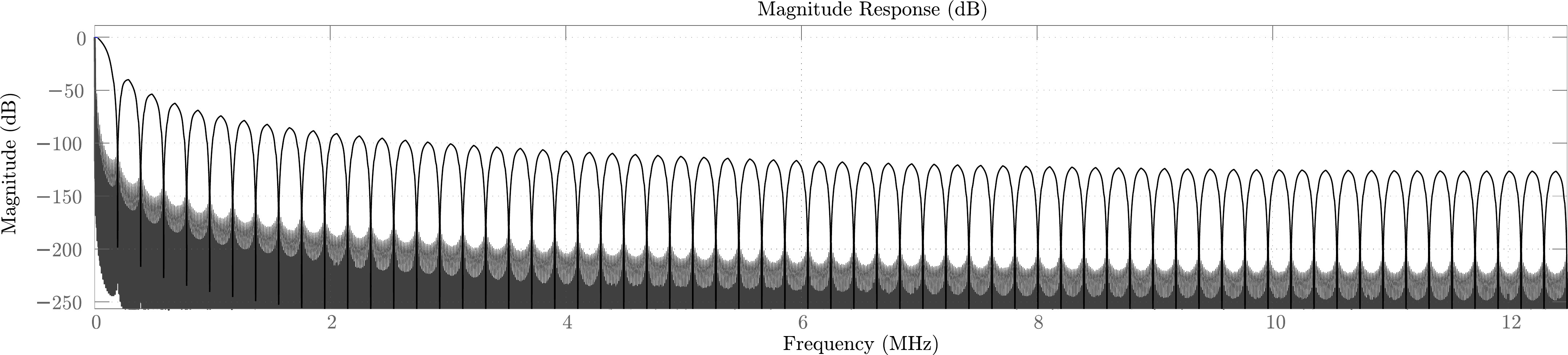}}
\caption{
	Filter response curves. The black (topmost) curve shows the frequency
	response of the CIC1 filter. The grey (middle) curve shows the
	combined response of the CIC1 and CIC2 filters. Finally, the solid
	grey (bottommost) region shows the combined responses of CIC1, CIC2,
	and FIR1.  The filters' noise performance, is dominated by aliasing,
	which depends on the width of the many nulls in this plot and is not
	evident here.  It, and the passband (which is too narrow to be visible
	in this figure) are examined in Fig.~\ref{fig:cic-passband3}.
	\label{fig:cic-passband}
}
\end{figure*}

\begin{figure*}
\centerline{\includegraphics[width=\textwidth]{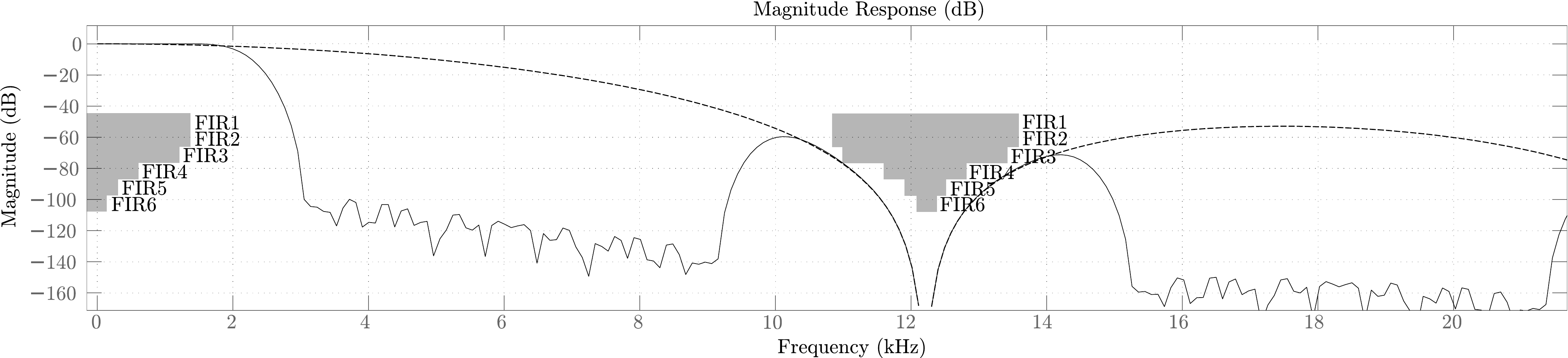}}
\caption{
	Spectral response of the filter chain after CIC2 (dashed; top) and
	after FIR1 (solid; bottom). The passband, shown on the left, has a
	ripple on the order of~$10^{-3}~\deci \bel$. The filter null shown
	near the center of the plot defines the alias-rejection performance of
	the chain, which exceeds $100~\deci \bel$ for the whole passband of
	FIR stages 4--6. The shaded regions show the passband width for each
	of the FIRs, any of which may be selected as demodulator outputs.
	\label{fig:cic-passband3}
}
\end{figure*}

\begin{figure*}
\centerline{\includegraphics[width=\textwidth]{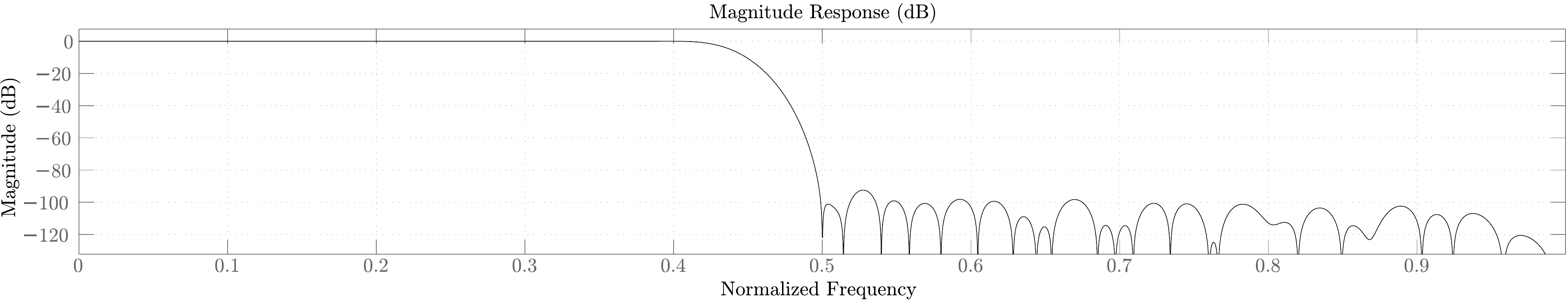}}
\caption{
	Normalized frequency response for FIR stages 2 through 6. Each of
	these filters decimates by a factor of 2; the passband covers 80\% of
	the full output bandwidth and exhibits ripple on the order
	of~$10^{-3}~\deci \bel$.
	\label{fig:stage23456}
}
\end{figure*}

The system's baseband spectral response is shown after CIC1, CIC2, and FIR1 in
Figures~\ref{fig:cic-passband}-\ref{fig:cic-passband3}. The spectral response
for filters FIR2 through FIR6, which are identical, is shown in
Fig.~\ref{fig:stage23456}.

\subsubsection{Buffering and MicroBlaze Interface}

The user may instruct the DFMD to send data directly from any of the FIR
stages to the output FIFO through a software-controlled multiplexer, such that
the data rate and signal bandwidth can be varied. This makes the system
immediately amenable to different detector bandwidths and to debugging and
measuring detector properties.  Each sample emerging from the selected FIR
stage is truncated to a 24-bit 2's-complement number and tagged with an 8-bit
channel identifier. The resulting 32-bit data is timestamped and written into
the data FIFO for retrieval by software.

We note that it is slightly redundant to store channel identifiers alongside
channel data. We used this approach since the data FIFO will overflow and lose
samples unless software is able to consume samples faster than they are
produced. While this is not the case under ordinary operation, the FIFO does fill
up during system boot or when data is not being streamed from the board; thus,
software must be able to detect and recover from disordered data.  This
is easy to accomplish with channel identifiers, and the additional FIFO
requirements are not onerous. (In future revisions, this problem may be
avoided by delivering samples via either DMA or a so-called packet FIFO.)

In the next section, we describe a timestamping process occurring in tandem
with the filtering described above.  After decimation and timestamping, the
aggregate data throughput is relatively low, and slight variations in signal
delay are no longer problematic. Thus, we drain the FIFO under software
control as described in Section~\ref{sec:streaming}.

\subsubsection{Timekeeping}

The DFMUX system is used in several experiments, each of which uses different
timekeeping standards and hardware. To integrate seamlessly with these
experiments, a timestamp multiplexer system accepts timestamps from different
timestamp units, including IRIG-B~\cite{IRIG1998} (derived from an external
GPS receiver) and an EBEX-specific~\cite{Oxley2004} timing system generated by
a separate timekeeping board.

Along with the format-specific timestamp decoders, we maintain local ``ticks''
counters. These counters are clocked by the DFMUX's oscillator and reset
whenever a new timestamp is decoded. They are used to increase the resolution
of timestamps available to the DFMUX within each interval between external
timestamps. They also permit the system to assign its own timestamps when no
external timestamps are provided to the board.

In both cases, decoded timestamps are less than 96 bits in size. These
timestamps fit into four 32-bit entries in the data FIFO, after reserving
space for 8-bit channel identifiers as described above for channel data.  Each
timestamp format is assigned four distinct channel identifiers.

Timestamps are added after multiplexing as follows: when the first channel is
emitted from the selected readout stage, the DMFD captures a timestamp of
user-specified format from the timestamp multiplexer.  After the last data
channel has been stored in the FIFO, this captured timestamp is broken into
four 24-bit numbers, tagged with the selected format's four assigned channel
identifiers, and placed into the FIFO.

Since timestamps are captured and associated with data after the decimation
process, the processing delays introduced by CICs and FIRs must be subtracted
from the timestamps to refer data back to the instant it were sampled. This
correction occurs in analysis software, since the timestamp format is opaque
to the DFMUX itself.

\section{Firmware and Software
	\label{sec:software}
}

From a user's perspective, the high-level control of the DFMUX is described
in~\cite{Macdermid2009}. In this section, we describe how the system is
designed to present a straightforward, flexible interface for real-time use
and automated control algorithms.

\subsection{Control}

Control instructions and queries are encoded as HTTP requests and received via
the board's Ethernet interface. They are decoded by a MicroBlaze processor
running the linux kernel. Providing the HTTP interface are two webservers:
a fully-featured, third-party webserver that serves a browser-accessible
interface, and a fastpath used only for rapid interaction with
hardware.

Although the HTTP interface allows all aspects of the DFMUX to be controlled,
it is a low-level interface that does not provide a method of interacting with
data from the signal path. To perform higher-level, possibly data-dependent
interactions (such as tuning the board), we provide a port of the Python
scripting environment, with a hardware-interface library that permits scripts
to control device registers directly. A Python library, pyWTL, supplies
higher-level interactions and allows board-level control to be performed in a
consistent, network-transparent manner, regardless of whether it runs on a
DFMUX or a PC.

Developing the higher-level code in an user-friendly environment like Python,
and allowing this code to run on either the DFMUX or a PC, permits physicists
as well as engineers to contribute to the development process.

\subsection{Data Streaming
	\label{sec:streaming}
}

In Sections~\ref{sec:signalpath} and~\ref{sec:software}, we described the
signal path and its control. In the following section, we discuss how data is
removed from the FIFO, packaged, and streamed across the network for storage.

Depending on which FIR stage has been selected, data from each channel is
streamed into the FIFO at a rate between $12~\kilo\hertz$ and $190~\hertz$,
for an aggregate throughput of between $13~\mathrm{ksps}$ and
$880~\mathrm{ksps}$ including timestamps. Data are removed from the FIFO in
\emph{frames} (sets containing one sample from each user-selected channel
and a single timestamp), and wrapped into packets containing several frames.
The number of frames per packet is chosen to maximize network throughput, i.e.
to maximize the size of each packet without causing fragmentation at the
network layer.

Once a packet is constructed, it is sent via unicast or multicast UDP to a
network address controlled by the Python or web interfaces. Since UDP does not
guarantee data transmission, network topology and hardware have been carefully
tested to ensure minimal packet loss. (Although TCP would permit
retransmission of dropped packets, a TCP-based transport would not permit
multicasting. It is also not clear, at expected data rates, that the
MicroBlaze and network hardware have the capacity to recover without further
packet losses.)

Packets are received by a companion utility (the ``parser'') and a Python
interface. The parser streams data to disk for both real-time monitoring
using KST~\cite{KST2009} and off-line analysis. Using SWIG~\cite{SWIG1997},
the Python interface integrates seamlessly with the pyWTL module described
above and is used for on-line tuning, control, and analysis tasks.

\section{System Performance}
\label{sec:performance}

\subsection{DMFS Noise}

There are three general classifications of noise in the DMFS: digital noise,
DAC noise, and analog noise. In this section, we quantify each of these noise
sources and determine the dominant source.

There are two sources of digital noise: that intrinsic to the DDS algorithm,
and quantization noise from the system's finite bit length. We model both of
these noises as additive uniform white noise, independent from other noise
sources. The noises thus add incoherently, and we may determine an overall
noise figure for the digital portion of the DMFS. Digital noise is summarized
in Table~\ref{tab:dmfs-noise-digital}. The overall RMS noise is approximately
1.16 LSBs at the DAC. This corresponds to a PSD of $2.5 \nano \volt/\sqrt
\hertz$ at the DAC.

% 1.16 LSB / 65536 * 0.5 Vpp / sqrt(12.5e6) = 2.5e-9

\begin{table}
\centerline{\begin{tabular}{r|c|c}
\bf{Noise Source} & \multicolumn{2}{c}{\bf{RMS Noise (DAC LSBs)}} \\
		& \bf{Per Channel} & \bf{Total} \\ \hline
DDS		& $32.5\times 10^{-3}$		& $130\times 10^{-3}$ \\
Truncation	& $289 \times 10^{-3}$		& $1.16$ \\ \hline
\bf{Total}	& $\mathbf{290 \times 10^{-3}}$	& $\mathbf{1.16}$ \\
\end{tabular}}
\vspace{0.5em}
\caption{
	Digital contributions to DMFS noise.
	\label{tab:dmfs-noise-digital}
}
\end{table}

The vendor documentation for the DAC suggests approximately $92 \deci \bel$
SFDR in our operating regime. If we model this as a white-noise floor, the
corresponding RMS noise is 1.65 DAC LSBs, slightly dominating the digital
noise. In terms of volts at the DAC, this corresponds to $3.6 \nano
\volt/\sqrt \hertz$.

% 1.65 LSB / 65536 * 0.5 Vpp / sqrt(12.5e6) = 3.6e-9

Noise in the DMFS's analog circuitry is dominated by its first-stage
amplifier, which has an input-referred noise of $2.5 \nano \volt/\sqrt
\hertz$.

Thus, the dominant source of noise in the modulator is the DAC itself,
followed by quantization noise in the digital portion of the system. The
overall noise is approximately $4.4 \nano \volt/\sqrt \hertz$ at the DAC.
Referred to an equivalent voltage at the DMFS' output (assuming an $100 \ohm$
load), the modulator's noise performance is summarized in
Table~\ref{tab:dmfs-noise}.

\begin{table}
\centerline{\begin{tabular}{c|c|c}
\multicolumn{2}{c|}{\bf{DMFS Gain}} & \bf{Noise} \\
\bf{Setting} & \bf{Gain} & ($\nano \volt/\sqrt \hertz$) \\ \hline
Low		& 0.48	& $2.1$ \\
Med. Low	& 1.09	& $4.7$ \\
Med. High	& 3.33	& $14.5$ \\
High		& 10.0	& $43.5$ \\
\end{tabular}}
\vspace{0.5em}
\caption{
	DMFS Noise, referred to the output assuming a 100-Ohm load.
	\label{tab:dmfs-noise}
}
\end{table}

\subsection{DMFD Noise}

In the DMFD, as in the DMFS, there are three general sources of noise: noise
in the analog circuitry, noise in the analog-to-digital converter (ADC), and
noise in the digital signal chain. In this section, we examine these noise
sources and determine the overall performance of the demodulator.

% Noise input at DMFD
%
% 1.3 nV/sqrt(Hz)	(at demod THS4131 input)
%	* [10000/10100 10000/2100 10000/300 10000/100]	(PGA)
%	* 2**14 / 2	(ADC 14 bits / 2 Vpp) = [	
%	1.0544158415841584e-05
%	5.0712380952380953e-05
%	0.0003549866666666667
%	0.00106496 ]
%
% At streamer:
%	* 0.6804	(mixer RMS)
%	* 8		(14 -> 17 bits)
%	/ 2		(17 -> 16 bits)
% = [	2.86953222e-05
%	1.38010836e-04
%	9.66075849e-04
%	2.89822755e-03 ]

Neglecting the source signal, noise in the analog signal chain is dominated by
the first amplification stage. These amplifiers have an input-referred noise
of $1.3~\nano \volt / \sqrt \hertz$.

The ADC exhibits a SNDR (signal to noise-plus-distortion ratio) of $73~\deci
\bel$, or $90~\nano \volt/\sqrt \hertz$ at the ADC ($734 \times 10^{-6}
\mathrm{ADC~ LSBs}/\sqrt \hertz$) at a sampling rate of 25 MSPS, assuming a
full-scale sinusoidal input signal.

% Calculation for the above:
% 2v full scale
%	* 1/sqrt(2)	(assume sinusoidal signal, full scale)
%	* 10^(-73/20)	(attenuate by 73 dB)
%	/ sqrt(12.5e6)	(convert to PSD)
%	= 90.0 nV/sqrt(hz)
%	* 2**14/2	(convert to ADC LSBs)
%	= 734 * 10^-6 ADC LSB/sqrt(hz)
%	* 0.6804	(mixer)
%	* 4		(convert to streamer LSBs)
%	= 0.0019965226027344181 streamer LSBs/sqrt(hz)

Noise in the digital chain is dominated by mixer effects and truncation. Each
of these noise sources are evaluated in this section.  (We neglect noise from
aliasing in the decimation filter chain, as it is subdominant.)

Because it is a nonlinear device, the mixer alters properties of noise in its
input. The mixer also contains a truncation stage, injecting noise of its own
after mixing.  As noted above, the mixer's performance depends on the
particular carrier frequency chosen, but its response to a Gaussian white
noise input with RMS amplitude $\sigma~\mathrm{ADC~LSBs}/\sqrt \hertz$ is
readily approximated as a white-noise output with RMS amplitude $0.6804
\sigma~\mathrm{ADC~LSBs}/\sqrt \hertz$. (The factor $0.6804$ is the RMS of the
mixer's coefficients; an ideal sinusoidal mixer would amplify noise by a
factor of $1/\sqrt{2}$) The distortion added by the mixer's truncation stage
may be modeled as additive uniform white noise of RMS amplitude $81.6\times
10^{-6}$ post-mixer LSBs/$\sqrt \hertz$ following the mixer, or $3.3 \times
10^{-3}$ streamer LSBs/$\sqrt \hertz$.

% Calculation for mixer noise:
% from pylab import *
% x = array([0,3,6,7,7,7,6,3]) / 8.0
% rms = sqrt(mean(x * x))
% Note: this is a gain applied to other noise sources, not its own noise
% source.

% Calculation for truncation noise:
% (1 LSB) / sqrt(12)	(RMS noise)
%	/ sqrt(12.5e6)	(to LSBs/sqrt(hz))
%	* 4		(to streamer LSBs)
%	= 0.00032659863237109049
%
% Referred to mezz inputs:
% (1 LSB) / sqrt(12)	(RMS noise)
%	/ sqrt(12.5e6)	(to post-mixer LSBs/sqrt(hz))
%	/ 0.6804	(to ADC LSBs/sqrt(hz))
%	* 2 / 2**14	(to ADC V/sqrt(hz))
%	/ [ gains ]
% = [	1.4796077863740684e-08
%	3.0764122290945978e-09
%	4.3948746129922819e-10
%	1.4649582043307607e-10 ]

To determine the overall noise level in the DMFD, we refer each source of
noise both back to the DMFD input (so it may be compared to physical signal
levels) and to the DMFD's streamed output (so it may be compared to
measurements.) These noise levels are summarized in
Table~\ref{tab:dmfd-noise}.

\begin{table}
\centerline{\begin{tabular}{l|c|c}
\multirow{2}{*}{\bf{Noise Source}} & \multicolumn{2}{c}{\bf{Intrinsic DMFD Noise}} \\ \cline{2-3}
 & $\volt/\sqrt \hertz$ & LSBs/$\sqrt \hertz$ \\ \hline
\multicolumn{3}{c}{\bf{Low Gain (0.99)}} \\ \hline
Analog	& $1.3 \times 10^{-9}$	& $2.9 \times 10^{-5}$ \\
ADC	& $9.0 \times 10^{-8}$	& $2.0 \times 10^{-3}$ \\
Digital	& $1.5 \times 10^{-8}$	& $3.3 \times 10^{-4}$ \\
\textbf{Total} & $\mathbf{9.2 \times 10^{-8}}$ & $\mathbf{2.0 \times 10^{-3}}$ \\ \hline
\multicolumn{3}{c}{\bf{Med. Low Gain (4.76)}} \\ \hline
Analog	& $1.3 \times 10^{-9}$	& $1.4 \times 10^{-4}$ \\
ADC	& $1.9 \times 10^{-8}$	& $2.0 \times 10^{-3}$ \\
Digital	& $3.1 \times 10^{-9}$	& $3.3 \times 10^{-4}$ \\
\textbf{Total} & $\mathbf{1.9 \times 10^{-8}}$ & $\mathbf{2.0 \times 10^{-3}}$ \\ \hline
\multicolumn{3}{c}{\bf{Med. High Gain (33.3)}} \\ \hline
Analog	& $1.3 \times 10^{-9}$	& $9.7 \times 10^{-4}$ \\
ADC	& $2.7 \times 10^{-9}$	& $2.0 \times 10^{-3}$ \\
Digital	& $4.4 \times 10^{-10}$	& $3.3 \times 10^{-4}$ \\
\textbf{Total} & $\mathbf{3.0 \times 10^{-9}}$ & $\mathbf{2.2 \times 10^{-3}}$ \\ \hline
\multicolumn{3}{c}{\bf{High Gain (100)}} \\ \hline
Analog	& $1.3 \times 10^{-9}$	& $2.9 \times 10^{-3}$ \\
ADC	& $9.0 \times 10^{-10}$	& $2.0 \times 10^{-3}$ \\
Digital	& $1.5 \times 10^{-10}$	& $3.3 \times 10^{-4}$ \\
\textbf{Total} & $\mathbf{1.6 \times 10^{-9}}$ & $\mathbf{3.5 \times 10^{-3}}$ \\
\end{tabular}}
\vspace{0.5em}
\caption{
	DMFD Noise Sources. Physical units are equivalent voltages at
	mezzanine inputs; numerical units are equivalent LSBs in the 16-bit
	values streamed across the network.
	\label{tab:dmfd-noise}
}
\end{table}

Once again, the use of truncation instead of convergent rounding
throughout the DMFD presents itself as a slight DC offset. For our
applications, the detectors are capable only of measuring the difference
signal between sky pointings, so no information is present at DC. Since any
offset is discarded during analysis, we make no attempt to correct for
additional bias introduced by truncation.

\subsection{Power Consumption}

Because its intended uses include balloon-based experiments such as EBEX, the
DFMUX's power consumption is a crucial performance measure. In EBEX, the DFMUX
is powered using (heavy) batteries. In addition, any heat generated by the
electronics must be removed from the gondola. This is a non-trivial challenge
task at high altitudes where the atmosphere is thin.

The DFMUX's power profile is shown in Table~\ref{tab:dfmux-power}. The first
row shows consumption for the DFMUX's digital circuitry only; the second
includes two analog mezzanines, which include the DACs, analog signal path,
and ADCs. Neglected in this power profile is the SQUID controllers, which
are separate boards that sit between the DFMUXes and the cryostat. (SQUID
controllers consume approximately $5~\watt$ each. One SQUID controller is
required for every 2 DFMUXes.)

\begin{table}
\centerline{\begin{tabular}{l|c|c|c}
\bf{Scenario} & \bf{Voltage} & \bf{Current} & \bf{Power} \\ \hline
Motherboard Only & $\pm~5.7~\volt$ & $1.58~\ampere $ & $9.01~\watt$ \\
With Mezzanines & $\pm~5.7~\volt$ & $2.63~\ampere $ & $15.0~\watt$ \\
\end{tabular}}
\vspace{0.5em}
\caption{
	Power consumption for the DFMUX.
	\label{tab:dfmux-power}
}
\end{table}

Because FPGAs are not ordinarily considered power-efficient, we contrast
the DFMUX's power dissipation per bolometer channel relative to the analog
fMUX, the mixed-signal implementation it supersedes.  The analog fMUX
\cite{Lanting2005,Dobbs2006,Carlstrom2009} is currently deployed at the South
Pole Telescope, and supports 840 electrically active bolometers with 60
boards. Each board includes 2 modules, each of which can support 7 bolometers.
Overall, SPT's analog fMUX crates draw $3~\kilo \watt$, for a power
consumption per bolometer of $3.6~\watt$. In a digital fMUX system using a
multiplexing factor of 16, power consumption per bolometer is
$250~\milli\watt$, for an overall power savings in excess of 90\%.

\subsection{FPGA Usage}

The FPGA resources used by the design are summarized in
Table~\ref{tab:dfmux-fpga}. The capacities listed are for the Virtex-4 LX160
FPGA, which is used on all but a handful of our boards. Resource utilization
is summarized for the signal path only, and for the entire design. The signal
path figures comprise the DMFS and DMFD, but neither the timestamp decoders
nor the FIFO connecting the DMFD to the processor bus. (The timestamp decoders
are relatively small; the FIFO consumes approximately 30 additional RAMB16s.)
The entire design includes the signal path, timestamp decoders, FIFOs,
MicroBlaze, network and memory interfaces, plus interfaces for the on- and
off-board peripherals and interfaces (voltage and temperature monitors, serial
ports, etc.) supported by the DFMUX.

\begin{table}
\centerline{\begin{tabular}{c|c|c|c}
\bf{Resource} & \bf{Number} & \bf{Device} & \bf{Percent} \\
\bf{Type} & \bf{Used} & \bf{Capacity} & \bf{Used} \\ \hline
\multicolumn{4}{c}{\bf{Signal Path Only}} \\ \hline
DSP48	& 33		& 96		& 34\% \\
RAMB16s	& 85		& 288		& 29\% \\
Slices	& 31,459	& 67,584	& 46\% \\ \hline
\multicolumn{4}{c}{\bf{Entire Design}} \\ \hline
DSP48	& 56		& 96		& 58\% \\
RAMB16s	& 273		& 288		& 94\% \\
Slices	& 48,023	& 67,584	& 71\% \\
\end{tabular}}
\vspace{0.5em}
\caption{
	FPGA Resource Consumption.
	\label{tab:dfmux-fpga}
}
\end{table}

\section{Discussion and Future Work
	\label{sec:conclusions}
}

We have described the signal path, control logic, and firmware of a
highly-integrated FDM system for TES bolometers. This system is physically
much smaller and less power-hungry than the mixed-signal system it replaces.

Because the bulk of the DFMUX's signal processing is performed on a FPGA which
may be easily reconfigured, the system has also found use in unexpected ways.
With only trivial hardware modifications, a DFMUX was used to measure the
surface accuracy of the South Pole Telescope's secondary
mirror~\cite{Padin2008}. A similarly modified DFMUX is used to measure the
angular position of a rotating optical element (half-wave plate) in the EBEX
telescope.

We are currently focusing on three improvements to the DFMUX. The first is an
attempt to reduce its power requirements (per bolometer channel) by improving
the design of its signal path.  The second goal is to increase the number of
bolometers multiplexed via each FDM module. We anticipate doubling or
quadrupling the total number of bolometers per DFMUX by simply scaling the
current design. Even greater channel densities may be achieved using polyphase
filter banks~\cite{Harris2004} or similar techniques. Finally, we are
investigating alternative biasing and feedback strategies that permit greater
increases in channel density.

\section{Acknowledgements}

This work was supported by the Natural Sciences and Engineering Research
Council of Canada (NSERC), the Canadian Institute for Advance Research
and the Canada Research Chairs program. We also gratefully acknowledge hardware
and software contributions from the Xilinx University Program. MD acknowledges
Sloan Fellowship funding.

\bibliographystyle{IEEEtran}
\bibliography{IEEEabrv,dfmux_paper}

\end{document}